# Fluctuation-Enhanced Sensing for Biological Agent Detection and Identification

## INVITED PAPER

Laszlo B. Kish, Hung C. Chang, Maria D. King, Chiman Kwan, James O. Jensen, Gabor Schmera, Janusz Smulko, Zoltan Gingl, and Claes G. Granqvist

*Abstract*—We survey and show our earlier results about three different ways of fluctuation-enhanced sensing of bio agent, the phage-based method for bacterium detection published earlier; sensing and evaluating the odors of microbes; and spectral and amplitude distribution analysis of noise in light scattering to identify spores based on their diffusion coefficient.

## I. General considerations

FLUCTUATON-enhanced sensing (FES) [1-24] separates, amplifies and analyzes the stochastic component of sensor signals in sensors where normally the steady-state or mean value of sensor signals are utilized only. The FES method for chemical agents was first patented about a decade ago in Sweden [1,2] (patents are in the public domain now) however that time the method was not yet called FES. Even before that, several groups reported the sensitivity of the conductance noise of different non-passivated resistors against the chemical environment and even mentioned their potential for applications [3,4]. The related Swedish patents were

Plenary Talk at the Nanoelectronic Devices for Defense and Security Conference, Fort Lauderdale, FL, October 1, 2009.

Manuscript received December 25,2009. This work was supported in part by the Army Research Office under contract W911NF-08-C-0031.

L.B. Kish (until 1999, Kiss) is with Texas A&M Univ., Dept. of Electrical and Computer Engineering, College Station, TX 77843-3128, USA (corresponding author, phone: 979-847-9071; fax: 979-845-6259; e-mail: Laszlo@ece.tamu.edu).

H.C. Chang, is with Texas A&M Univ., Dept. of Electrical and Computer Engineering, College Station, TX 77843-3128, USA (e-mail: hungchih@neo.tamu.edu).

M.D. King is with Texas A&M Univ., Dept. of Mechanical Engineering, College Station, TX 77843 (email: mdking@neo.tamu.edu).

C. Kwan is with Signal Processing, Inc., Rockville, MD 20850 USA (email: chiman.kwan@signalpro.net).

J.O. Jensen is with the US Army, RDECOM, ECBC, Edgewood, MD 21040 USA (email: jim.jensen@us.army.mil).

G. Schmera is with the Space and Naval Warfare Systems Center, San Diego, CA 92152 USA

J. Smulko is with Gdansk University of Technology, WETiI, ul. G. Narutowicza 11/12, 80-952 Gdansk, Poland (email: jsmulko@eti.pg.gda.pl).

Z. Gingl is with University of Szeged, Dept. of Experimental Physics, Dom ter 9, Szeged, H-6720, Hungary (email: gingl@physx.u-szeged.hu).

C.G. Granqvist is with Uppsala University, Angstrom Laboratory, P.O. Box 534, SE-75121, Uppsala, Sweden (email: Claes-Goran.Granqvist@angstrom.uu.se).

authored because a mathematical analysis and related experimental feasibility studies with various commercial sensors (published later [5-7]) indicated a potential for FES. Today's name, Fluctuation-Enhanced Sensing, was given by John Audia (then at SPAWAR, San Diego) in 2001 during a visit to TAMU (College Station). Even though, FES was first introduced for gas sensing [1-3,5-7], the principle allows virtually the sensing of any chemical or physical agent. In the case of gas sensing, the related interactions at the nanoscale provide specific fluctuations of the sensor signal with characteristic stochastic dynamics that provides enhanced sensitivity and sensory information.

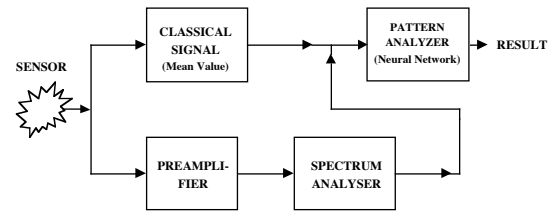

Fig. 1. Fluctuation-enhanced sensing scheme with power density spectra [5,6].

After a decade-long study of fluctuation-enhanced sensing (e.g. [5-20]) and limited body of literature of theoretical studies [13,22-24], enough experience has been acquired to confidently make the following basic claims:

i) FES is not a miracle-tool of sensing. However, it has a great potential, and it often provides unique results or it works at conditions where the corresponding classical sensing scheme cannot be used.

ii) Rule of thumb: The usual sensor signal, which is the average value of a physical quantity, has always less information than the spontaneous fluctuations of that physical quantity: the FES signal.

iii) FES always requires more efforts than the corresponding classical sensing. Minimal pre-requirements are significant AC pre-amplification and the isolation of relevant external disturbances.

iv) The smaller the characteristic length of sensor the higher

is the sensitivity and the greater is the sensory information. Nanostructures as sensors will have great potential for commercial applications as soon as they can be fabricated with sufficient stability and reproducibility.

In the present paper we briefly survey our results related to the sensing of bacteria and bacterial spores by FES. The two bacterium sensing/detection methods are demonstrated by experiments and they have a strongly empirical nature. The bacterium spore detection scheme is theoretical with firm base and they are confirmed by computer simulations.

## II. BACTERIUM IDENTIFICATION BY FES AND PHAGES

We have already surveyed this topic at several forums and here we mention it only for the completeness of this survey for biological sensing utilizing FES. In 2005, a FES-based method for prompt detection and identification of bacteria was proposed [18-20]. The method, SEnsing of Phage-Triggered Ion Cascade (SEPTIC), is detecting and analyzing the electrical field caused by the stochastic emission of ions during phage infection. Here "ions" is a term of biochemists and it simply means the mixtures of various ionized salts dissolved in water where the charge balance of positive and negative ions is not known. The infected bacteria emit about $10^8$ ions into the ambient fluid and the detection and identification is carried out by measuring and analyzing the voltage fluctuations between two metallic electrodes. The sizes and the distance of electrodes are in the order of a micron. There is a small DC current flowing between the electrodes and that collects infected bacteria at the electrodes. The system is a concentration cell (battery) with fluctuation ion concentration/gradients and the result is obtained in a few minutes after mixing the bacterium and corresponding phage solutions. In the case of phage infection the power density spectrum grows and it will follow a $1/f^2$ asymptotic shape, while without reaction (phage infection), it is weaker and shows the well-known $1/f$ shape, see Fig. 2. Because the phage infection is a very selective process, where only bacteria of a specific strain are infected, a futuristic SEPTIC-biochip containing an array of sensors, where each sensor is sensitized with a different phage, would be able to detect and identify a library of bacteria with high speed and selectivity, see Figure 3.

There are many important unsolved problems about the SEPTIC method [19-20]. Some examples follow here:

1. What is the optimal electrical field and geometry?

2. Effect of the salt concentration and practical biological ambient (blood, etc.) ?

3. Response of other phages and bacteria?

4. The timing, dynamics and ion composition of ion emission?

5. How to construct a pen-size biolab with a disposable biochip for the instantaneous detection and identification of a library of bacteria?

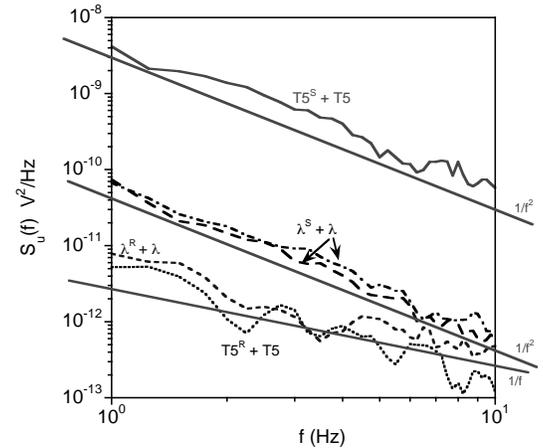

Fig. 2. Phage T5 infection of bacteria *E.coli* T5$^S$ or *E.coli* T5$^R$. Phage $\lambda$ infection of bacteria *E.coli* $\lambda^S$ or *E.coli* $\lambda^R$. The phage $\lambda$ is a genetically modified phage with reduced reactibility; this is why the SEPTIC signal is weaker. Control experiments with phage Ur-$\lambda$ and bacteria *E.coli* $\lambda^R$, and with phage T5 and bacteria *E.coli* T5$^R$. The fluctuations were recorded for 2 minutes.

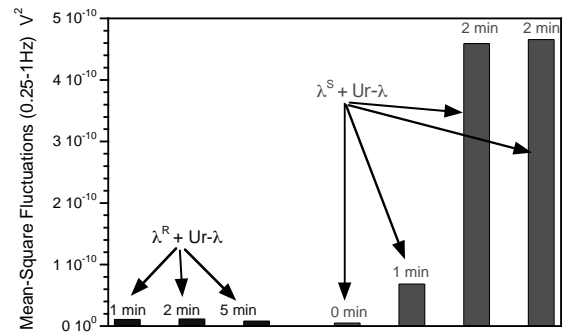

Fig. 3. Right columns: total mean-square fluctuations in the frequency range of 1-10Hz during the infection of bacteria *E.coli* $\lambda^S$ by phages Ur-$\lambda$ after various incubation times. The two 2-minutes experiments show the reproducibility of the effect. Left columns: the same experiments with resistant bacteria *E.coli* $\lambda^R$.

Finally, we note that the successful continuation of such developments require a strongly interdisciplinary effort where the list of required expertise include: chip engineering, phase biology, electrochemistry and signal processing.

## III. FLUCTUATION-ENHANCED SENSING OF ODORS OF MICROBES

Very recently [8], we had studied the specificity of power density spectra measured in the resistance fluctuation of commercial Taguchi sensors exposed to bacterial odors. Later, in order to provide a new type of pattern recognition method with ultra-low power consumption, while using FES with room-temperature sensors, we developed and tested a simple way to generate binary patterns based on spectral slopes in different frequency ranges [21]. First, the deviation between the local slope (average slope in a given frequency range) and the global slope of the power density spectra of FES signals is calculated. Then the sign of this deviation is evaluated for each non-overlapping frequency ranges and used as a pattern. Such patterns can be considered as binary "fingerprints" of odors. The feasibility of the new method has experimentally been demonstrated with a commercial semiconducting metal oxide (Taguchi) sensor exposed to bacterial odors (vegetative Escherichia coli, and Anthrax-surrogate Bacillus subtilis spores) and processing their stochastic signals. With a single Taguchi sensor, the situations of empty chamber, tryptic soy agar (TSA) medium, or TSA with bacteria could be distinguished with 100% reproducibility. The bacterium numbers were in the range of $2.5*10^4 - 10^6$. Albeit, these numbers were not low, this study was performed only to demonstrate how much improvement FES can yield, especially that, without this method, the given Taguchi sensors provided zero information about the bacteria. Examples of the measured spectra of (normalized) resistance-fluctuation are shown in Figs. 4-6 and the binary pattern generated by the new method is shown in Fig 7.

To illustrate the relevance for ultra-low power consumption, we showed that this new type of signal processing and pattern recognition task can be implemented by a simple analog circuitry and a few logic gates with total power consumption in the microWatts range. Fig. 8 shows the simplicity of the pattern recognition unit. No computers, microprocessors or extensive calculations were needed, just elementary logic decisions. Taguchi sensors were used as a demonstration only and, for a low-power application, room-temperature nanoparticle film sensors are envisioned (as soon as they will become commercially available) with this FES scheme.

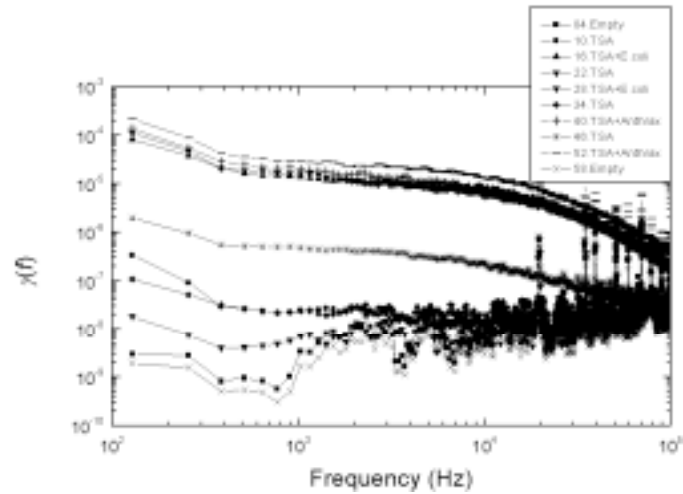

Fig. 6. Repeated and expanded experiments with new samples (compare with Fig. 4).

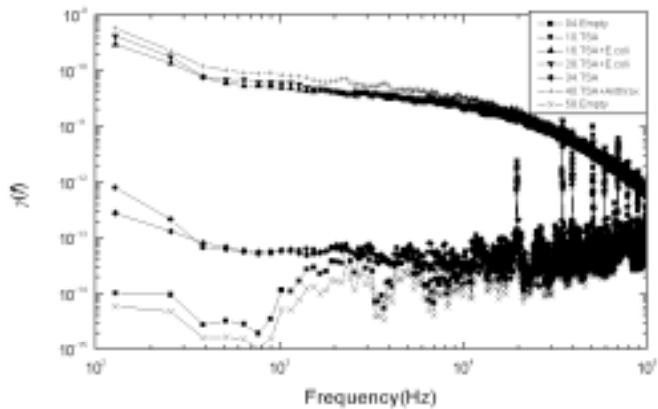

Fig. 4. Normalized power density spectra of the resistance fluctuations of the sensor SP32 measured in the sampling-and-hold [6] working mode. Each sample had one million bacteria. The alias "Anthrax" stands for Anthrax surrogate Bacillus subtilis.

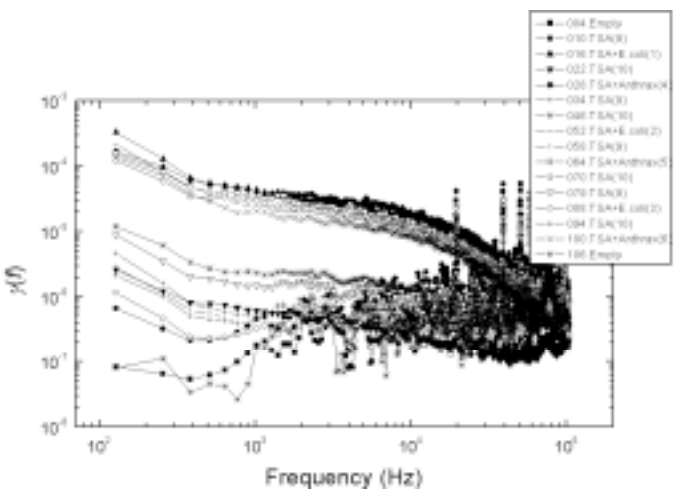

Fig. 5. Repeated and expanded experiments with new samples (compare with Fig. 4).

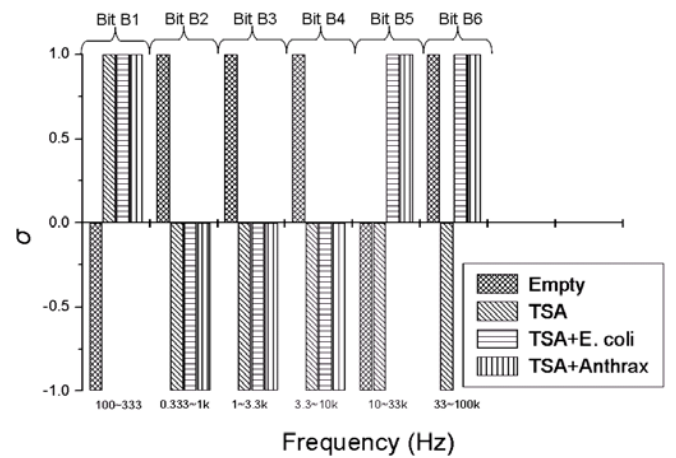

Figure 7. The spectra in Figures 4-6 yield the same 6-bits pattern. However, gradually decreasing the bacterium number to 25 thousands, indicated that bit B5 was not reproducible.

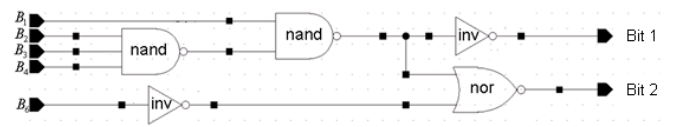

Figure 8. The Boolean logic circuit to realize the binary pattern recognition for the binary patterns shown in Figure 7. Bit B5 was skipped thus no input line $B_5$ is used.

## IV. SPORE DETECTION BY SPECTRAL ANALYSIS OF NOISE IN LIGHT SCATTERING

Bacterial spores in fluid execute Brownian motion. If the fluid is exposed to a narrow laser beam, the reflected/scattered light will execute a diffusion noise phenomenon because the number of spore will fluctuate in the beam. Randomly leaving a returning spores will contribute to the long-term correlations in the observed fluctuations. Photon correlation spectroscopy [25] and fluorescence correlation spectroscopy [26] (these are old photonics-based predecessors of FES) are observing the autocorrelation function of such fluctuations caused by dispersed particles in fluid and provides the value of the diffusion coefficient of the particle.

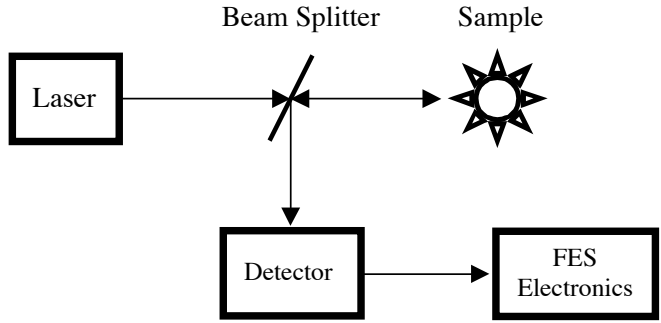

Fig. 9. Outline of the optical setup to study diffusion noise in light scattering of spores dispersed in fluid.

Our goal is to enhance this light scattering method in both the time and the frequency domains, see Fig. 9. During the recent years, we have been developing FES-based analysis tools for [10,17] similar diffusion problems. The observed spectra in the high-frequency limit show the so-called universal frequency scaling of diffusion noise which is proportional to $f^{-1.5}$. In the low frequency limit, the spectrum approaches white noise (in 2 or more dimensions and/or with finite-size diffusion volume). The characteristic crossover frequency between white and the $f^{-1.5}$ range is related to the reciprocal of the diffusion time through the beam cross-section. Exact spectra can be obtained by computer simulations and they can be matched by the measured spectrum to determine the diffusion coefficient. In an idealistic situation, the spore can be identified from its diffusion coefficient.

The main advantage of being in the frequency domain instead of the using autocorrelation functions is that we can apply advanced spectral tools, such as cross spectra with two adjacent light beam to reduce background noise; and bispectrum to obtain information related to the non-Gaussianity of the fluctuations [10]. A demonstration of the enhanced sensing information is shown in Figure 10. The presence of different types of particles in the mixture of three different particles was estimated from these spectra. Whenever enough data were available, the bispectra worked better [10].

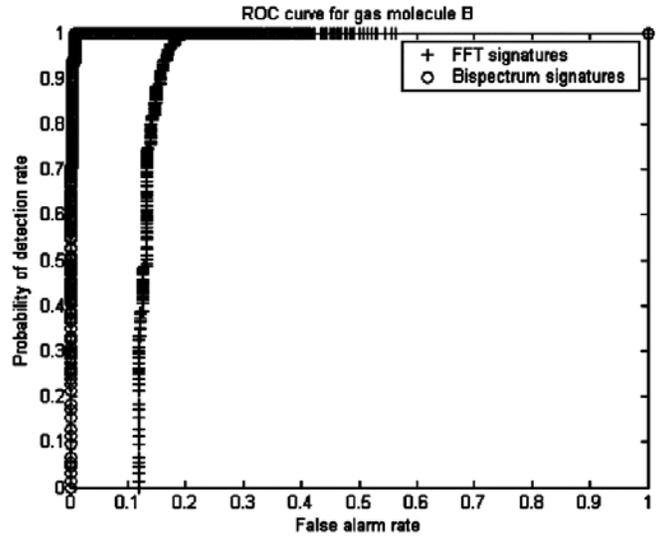

Fig. 10. Receiver Operation Characteristic (ROC) curves with power density spectra and bi-spectra.

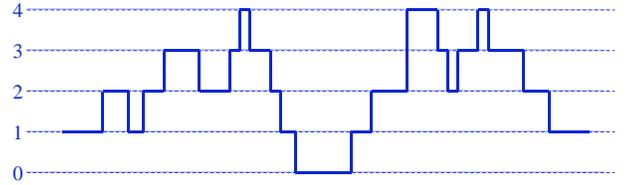

Fig. 11. Diffusion noise signal with 4 particles diffusing between two subspaces. The sensor produces a signal proportional to the number of particles in one of the subspaces.

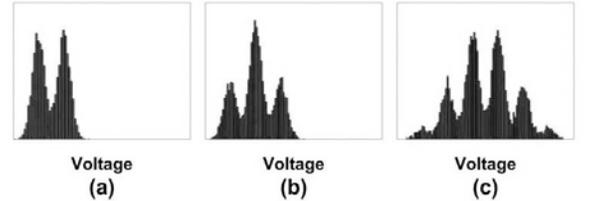

Fig. 12. Amplitude distribution function of the diffusion noise with 2, 3 and 6 particles in the presence of background noise.

Can we use the method in a small system with a few particles? The answer is yes; moreover, the smaller the system is the better the theoretical situation for the FES provided measurement duration is not a problem. Here we show an example with the amplitude distribution function that is usually Gaussian in large sensors with many particles. However, for a small sensor and/or low number of particles, it will be a non-Gaussian function (binomial distribution), see Figs. 11 and 12. The number of spores in the system is equal to the number of peaks of the amplitude distribution function. This methods serves a way for calibration by first principles [17].

Finally, applying techniques in the amplitude domain such as binomial distribution analysis for particle counting or zero-crossing statistics requires few calculations and very low power consumption [10].

## V. CONCLUSION

Utilizing the stochastic component of the signal of bio-sensors can significantly enhance sensory information and can provide other advantages, too, such as reduced power need.